\documentclass[useAMS,usenatbib,a4paper]{mn2e}

\usepackage{graphicx}
\usepackage{txfonts}
\usepackage{url}

\begin{document}

\title{On the robustness of the ammonia thermometer}

\author[S. Maret, A. Faure, E. Scifoni and
  L. Wiesenfeld]{S. Maret$^{1}$, A. Faure$^{1}$,
  E. Scifoni$^{1, 2}$ and L. Wiesenfeld$^{1}$\\
  $^{1}$Laboratoire d'Astrophysique de Grenoble, Observatoire de
  Grenoble, Universit\'e Joseph Fourier, CNRS, UMR 5571 Grenoble,
  France\\
  $^{2}$Frankfurt Institute for Advanced Studies, Johann Wolfgang
  Goethe University, Frankfurt am Main, Germany \\}

\date{Accepted 2009 June 22. Received 2009 June 22; in original form 2009 April 30}

\newcommand{\mnras}{MNRAS}
\newcommand{\nat}{Nature}
\newcommand{\aap}{A\&A}
\newcommand{\apj}{ApJ}
\newcommand{\apjs}{ApJSS}
\newcommand{\apjl}{ApJ}
\newcommand{\jcp}{J. Chem. Phys.}
\newcommand{\araa}{ARA\&A}

\pagerange{\pageref{firstpage}--\pageref{lastpage}} \pubyear{2009}

\maketitle

\label{firstpage}

\begin{abstract}
  Ammonia inversion lines are often used as probes of the physical
  conditions in the dense ISM. The excitation temperature between the
  first two para metastable (rotational) levels is an excellent probe
  of the gas kinetic temperature. However, the calibration of this
  ammonia thermometer depends on the accuracy of the collisional rates
  with H$_{2}$. Here we present new collisional rates for
  ortho-NH$_{3}$ and para-NH$_{3}$ colliding with para-H$_{2}$ ($J=0$)
  and we investigate the effects of these new rates on the excitation
  of ammonia.  Scattering calculations employ a new, high accuracy,
  potential energy surface computed at the coupled-cluster CCSD(T)
  level with a basis set extrapolation procedure. Rates are obtained
  for all transitions involving ammonia levels with $J\leq 3$ and for
  kinetic temperatures in the range 5$-$100~K. We find that the
  calibration curve of the ammonia thermometer -- which relates the
  observed excitation temperature between the first two para
  metastable levels to the gas kinetic temperature -- does not change
  significantly when these new rates are used.  Thus, the calibration
  of ammonia thermometer appears to be robust.  Effects of the new
  rates on the excitation temperature of inversion and
  rotation-inversion transitions are also found to be small.
\end{abstract} 


\section{Introduction}

Since its discovery in the interstellar medium forty years ago by
\cite{Cheung69a}, ammonia has been widely used as a probe of the
physical conditions in a variety of interstellar environments, ranging
from pre-stellar cores, molecular clouds, to external galaxies
\citep[see][for a review]{Ho83}. The peculiar structure of the
molecule makes ammonia lines excellent tracers of the density and
temperature in these environments. NH$_{3}$ is a symmetric top
molecule, whose rotational levels can be denoted by two quantum
numbers, the total angular momentum $J$, and its projection $K$ along
the $C_{3v}$ molecular axis. Owing to the possible relative
orientations of the hydrogen spins, two distinct species exists:
ortho-NH$_3$ ($K=3n$, with $n$ an integer; hereafter o-NH$_3$) and
para-NH$_3$ ($K\ne 3n$; hereafter p-NH$_3$).  As both radiative and
non-reactive collisional transitions do not change the spin
orientations, transitions between o-NH$_3$ and p-NH$_3$ are forbidden.
Each of the rotational energy levels (with the exception of those with
$K=0$) is further split into two sub-levels which can be denoted
either by the inversion symmetry of the vibrational wave functions or
by the symmetry index $\epsilon=\pm 1$ \citep[see Fig.~1 and Eq.~29
in][] {rist93}\footnote{In this paper, we denote each level by its
  symmetry index $\epsilon$, with a $+$ sign for $\epsilon=+1$ and a
  $-$ sign for $\epsilon=-1$. We refer the reader to the Fig.~1 of
  \citet{rist93} for an energy diagram of the molecule.}. This
splitting is caused by the inversion motion of the molecule, and the
corresponding inversion transitions fall in the range $\lambda \sim
1$~cm. Electric dipole transitions rules ($\Delta J=0, \pm 1, \Delta
K=0$) prevent radiative transitions between different $K$ ladders to
occur. Consequently, the lowest inversion doublets in each $K$ ladder
(i.e.  with $J = K$) are metastable; they can be relaxed only through
collisions.

For this reason, the relative population of the first two metastable
inversion doublets of p-NH$_3$, $J_{K,\epsilon} = 1_{1,\pm}$ and
$2_{2,\pm}$, depends solely on the kinetic temperature. Considering
only the first three doublets, $1_{1,\pm}$, $2_{2,\pm}$ and
$2_{1,\pm}$, and assuming that the population of the $2_{1,\pm}$
doublet is much smaller than that of the $2_{2,\pm}$,
\citet{Walmsley83} showed that the excitation temperature between the
two lowest doublets is given by the analytic formula:

\begin{equation}
  T_{1,2}^A = T \left[ 1 + \frac{T}{T_{0}} \, 
    \mathrm{ln} \, \left( 1 + \frac{C \left( 2_{2} \rightarrow 2_{1} \right)}
    {C \left( 2_{2} \rightarrow 1_{1} \right) } \right) \right]^{-1}
  \label{eq:1}
\end{equation}

\noindent where $T_{0}$ is the energy difference between the first two
metastable doublets ($\sim 41.7$~K), $T$ is the kinetic temperature,
$C \left( 2_{2} \rightarrow 2_{1} \right)$ is the collisional
excitation rate (averaged over the symmetry index $\epsilon$) between
the $J_K = 2_2$ and $2_1$ rotational levels, and $C \left( 2_{2}
  \rightarrow 1_{1} \right)$ is the collisional de-excitation rate
between the $2_2$ and $1_1$ levels. Thus, if one knows the $T_{1,2}$
excitation temperature, one can derive the kinetic temperature of the
gas, effectively using ammonia inversion lines as a ``thermometer''.

Observationally, $T_{1,2}$ can be determined by observing the
hyperfine components of the $1_{1,-} \rightarrow 1_{1,+}$ and $2_{2,+}
\rightarrow 2_{2,-}$ inversion transitions.  The inversion doublets
have indeed hyperfine components, which are due to the interaction
between the electric quadrupole moment of the N nucleus and the
electric field gradient created by the electrons. If one assumes that
the excitation temperature of each hyperfine components (within a
given rotational state) is the same, then one can derive the opacity
$\tau(1_1)$ and $\tau(2_2)$ of the $1_1$ and $2_2$ multiplets from the
relative intensity of each hyperfine component \citep{Barrett77}.
$T_{1,2}$ is then calculated from the following formulae
\citep{Ho79,Hotzel02}:

\begin{equation}
  T_{1,2} = -T_{0} \big/ \ln \left( \frac{9}{20}
      \frac {\tau(2_2)}{\tau(1_1)} \right)
\label{eq:2}
\end{equation}

In order to ``calibrate'' the ammonia thermometer, i.e. to compute the
kinetic temperature from the measured $T_{1,2}$ excitation
temperature, a good knowledge of the collisional rates of NH$_{3}$
colliding with H$_2$ is necessary (see Eq. \ref{eq:1}).  Although a
large number of measurements have been made on the NH$_3-$H$_2$
system, using in particular double resonance \citep[e.g.][]{daly70},
crossed beam \citep[e.g.][]{schleipen93} and pressure broadening
\citep[e.g.][]{willey02} experiments, laboratory data generally do not
directly provide state-to-state rate coefficients. As a result,
radiative transfer models can exclusively rely on theoretical
estimates. On the other hand, laboratory measurements are crucial to
establish the predictive abilities of theory and, in particular, of
the potential energy surfaces (PES).

Following the pioneering work of \cite{morris73}, \cite{Walmsley83}
performed statistical equilibrium calculations based on the
theoretical NH$_3-$He collisional rates of \cite{Green80}.
\cite{Danby88} then determined more accurate collisional rates for
collisions of NH$_3$ with p-H$_{2}$($J=0$) and used them to
recalibrate the ammonia thermometer.  The scattering calculations of
\cite{Danby88} were based on the \emph{ab initio} NH$_3-$H$_2$ PES of
\citet{danby86}. This latter was subsequently improved to investigate
propensity rules at selected collisional energies for ammonia
colliding with both p- and o-H$_2$ \citep{offer90,rist93}. More
recently, high accuracy \emph{ab initio} calculations have been
performed by \citet{mladenovic08} to explore the topographical
features of the NH$_3-$H$_2$ interaction. In the present paper, we
present new collisional rates based on the determination of a new,
highly accurate, NH$_3-$H$_2$ \emph{ab initio} PES.  Ammonia and
hydrogen molecules are treated as rigid rotors and H$_2$ is further
constrained in the scattering calculations to be in the (spherically
symmetrical) para-$J=0$ state, as in \citet{Danby88}.  Hence, the main
difference between the present collisional rates and those of
\citet{Danby88} arises from the PES. These new rates are then used to
estimate the robustness of the ammonia thermometer and the excitation
of ammonia lines under the condition that prevails in cold molecular
clouds, prestellar cores and protostars ($T <$ 100~K). For this, we
compute the excitation of both o-NH$_{3}$ and p-NH$_{3}$ using a
non-LTE radiative transfer code. The paper is organized as follows. In
\S2 we present both the new NH$_3-$H$_2$ PES and the scattering
calculations. Our excitation computations and the comparison with
earlier computations are presented in \S~\ref{sec:non-lte-excitation},
and \S \ref{sec:conclusions} concludes this article.

\section{Potential energy surface and collisional rates}
\label{sec:potent-energy-surf}

\subsection{Potential Energy Surface}
\label{abinitio}

The ammonia and hydrogen molecules were both assumed to be rigid. This
assumption is adequate here because {\it i)} the investigated
collisional energies are below the first vibrational excitation
threshold of ammonia \citep[$\nu_2=932.4\;
\mathrm{cm}^{-1}$][]{rajamaki04} and {\it ii)} the corresponding
collision time scales are much faster than the inversion motion of
ammonia. Monomer geometries were taken at their ground-state average
values, as recommended by \citet{faure05}. The average structure of
NH$_3$ was derived from the high-accuracy calculations of
\citet{rajamaki04}: $r_{\rm NH}=1.9512\,\mathrm{Bohrs}$, and
$\widehat{\mathrm{HNH}}=107.38^{\circ}$.  The ground-state average
geometry of H$_2$ was taken as $r_{\rm HH}=1.4488\, \mathrm{Bohrs}$
\citep[e.g.][]{faure05}. The conventions of \citet{phillips94} were
employed in defining the NH$_3$-H$_2$ rigid-rotor (five-dimensional)
coordinate system (one H atom lies in the $(x, z)$ plane).

The NH$_3-$H$_2$ PES was constructed using the following two step
procedure: {\it (i)} a reference PES was computed from a large set
(89,000 points) of CCSD(T)\footnote{CCSD(T) stands for the coupled
  cluster method with noniterative evaluation of triple excitations.}
calculations using the Dunning's correlation consistent aug-cc-pVDZ
basis set; {\it (ii)} this reference surface was calibrated using a
complete basis set (CBS) extrapolation procedure based on a smaller
set (29,000 points) of CCSD(T)/aug-cc-pVTZ calculations. A CBS-type
extrapolation was applied to the correlation part of the interaction
energy and was performed using a two-point $X^{-3}$ type
extrapolation, where $X$ is the cardinal number corresponding to the
basis set, as described in \citet{jankowski05}. The self-consistent
field (SCF) contribution was not extrapolated but was taken at the
aug-cc-pVTZ level. All basis sets were supplemented with midbond
functions and all calculations were counterpoise corrected as in
\citet{jankowski05}. The same strategy was recently applied to
H$_2$CO$-$H$_2$ \citep{troscompt09}.

Grid points were chosen for 29 fixed intermolecular distances $R$ (in
the range $3-15$~a$_0$) {\it via} random sampling for the angular
coordinates of H$_2$ relative to NH$_3$. At each intermolecular
distance, the interaction energy was then least square fitted using a
120 terms expansion for the angular coordinates, using Eq.~(3) of
\citet{phillips94} adapted to the C$_{3v}$ symmetry of NH$_3$. This
expansion includes anisotropies up to $l_1=11$ and $l_2=4$, where the
integer indices $l_1$ and $l_2$ refer to the tensor ranks of the
angular dependence of the NH$_3$ and H$_2$ orientation, respectively.
The CBS correction was fitted over a subset of only 46 angular terms
with $l_1\leq 7$ and $l_2\leq 4$. We note that the expansion
restricted to p-H$_2$ ($J=0$), in which all terms with $l_2\ne 0$ are
eliminated, includes only 24 terms. The accuracy of the angular
expansions was monitored using a self-consistent Monte Carlo error
estimator. A cubic spline interpolation was finally employed over the
whole intermolecular distance range and was smoothly connected with
standard extrapolations to provide continuous radial expansion
coefficients suitable for scattering calculations. Technical details
on the fitting strategy can be found in \citet{valiron08}. The
accuracy of the final five-dimensional fit was found to be better than
1~cm$^{-1}$ in the long-range and minimum region of the interaction
($R>5$~Bohrs). The accuracy of the above procedure was also checked
against a moderate set (1,200 points) of ``high-cost'' CCSD(T)-R12
calculations which offer a direct way of reaching the basis set limit
value within a single calculation, that is without extrapolation
\citep{noga94}. The RMS error between the final fit and the benchmark
CCSD(T)-R12 values was found to be lower than 1~cm$^{-1}$ in the whole
attractive part of the interaction ($R\geq 6$~Bohrs). We emphasize
that the intrinsic accuracy of CCSD(T) calculations at the basis set
limit is $\sim 1$~cm$^{-1}$.

Constraining H$_2$ in its lowest para level ($J=0$) is strictly
equivalent to averaging the PES over the H$_{2}$ rotational motion.
The global minimum of this averaged PES lies at -85.7~cm$^{-1}$ for
$R$=6.3~Bohrs, with H$_2$ in an almost equatorial location,
equidistant from the two closest H atoms of ammonia. A similar
location was found for the global minimum of the NH$_3-$He
interaction, but with a significantly more shallow potential well at
$\sim -33$~cm$^{-1}$ \citep{hodges01}. The five-dimensional PES,
including the anisotropy of H$_2$, is of course qualitatively
different: the global minimum, as deduced from our fit, lies at
$-267$~cm$^{-1}$ for $R=6.1$~Bohrs, with H$_2$ colinear with the
C$_{3v}$ axis of ammonia at the nitrogen end. It is interesting to
compare this result with the recent calculations of
\citet{mladenovic08}: these authors found the global minimum of the
NH$_3-$H$_2$ interaction at a similar location with a comparable,
although significantly smaller, binding energy ($-253$~cm$^{-1}$). As
their calculations were performed at a similar level of accuracy
(CCSD(T) method and aug-cc-pVQZ basis sets), this difference most
likely reflects monomer geometry effects. Detailed comparisons will be
investigated in dedicated future works.

\begin{figure}
  \centering 
  \includegraphics*[width=\columnwidth]{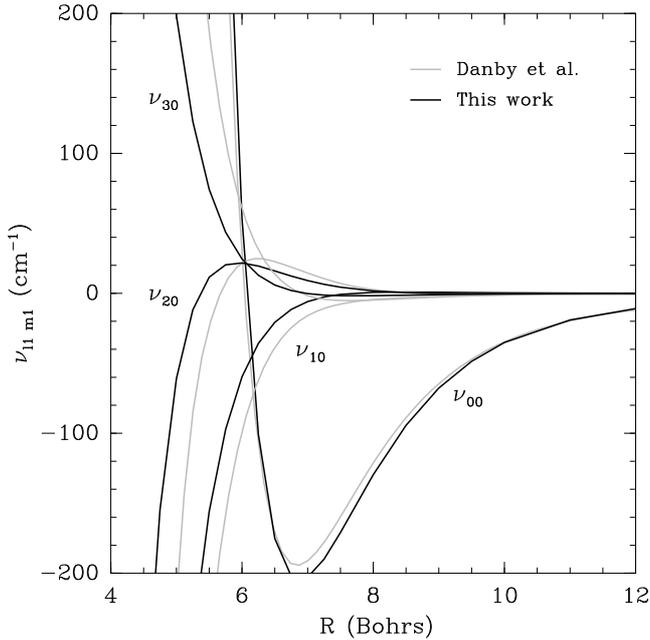}
  \caption{Comparison between the first four angular expansion
    coefficients, $v_{l_{1} m_{1}}$, of the NH$_3-$H$_2$ PES
    \citet{danby86} with those from the present work, as a function of
    the intermolecular distance $R$.}
  \label{fig:expcoef}
\end{figure}

Now, in order to compare the present NH$_3-$H$_2$ PES with that
employed by \citet{Danby88}, we present in Fig.~\ref{fig:expcoef} a
comparison of the angular expansion coefficients $v_{l_{1} m_{1}}$.
The definition of these coefficients is given in Eq.~(2.1) of
\citet{danby86} and their values are listed in their
Table~4\footnote{$v_{l_{1} m_{1}}$ corresponds to $v_{\lambda \mu}$ in
  Eq.~(2.1) of \citet{danby86}.}. Only the first four are plotted for
clarity in the figure. Despite significant differences at short-range
($R\leq 6$~Bohrs), the overall agreement between the two sets of
coefficients is quite reasonable. This comparison {\it i)} indicates
the good quality of the \emph{ab initio} calculations of
\cite{danby86} and {\it ii)} suggests moderate effects of the new PES
on the dynamics, as shown below.

Finally, it should be noted that the PES of \citet{danby86} has been
previously checked against laboratory measurements:
\citet{schleipen93} and \citet{willey02} reported, respectively,
symmetry-resolved state-to-state and broadening cross sections. In both
cases, a good overall agreement was obtained between theory and
experiment, suggesting the adequacy of the PES. Discrepancies were
however noted, in particular strong propensity rules predicted by
theory for NH$_3-$p-H$_2$ were observed in experiment to a much
lower extent (see below).

\subsection{Scattering calculations}
\label{scatt}

The quantal theory for scattering of a symmetric top with an atom or a
structureless molecule like H$_2$~($J$=0) can be found in
\citet{green76}. The extension of the formalism to the scattering of a
symmetric top with a linear molecule can be found in \citet{offer90}
and \citet{rist93}.  In the present work, calculations were performed
using the (nonreactive) scattering code \textsc{MOLSCAT}
\citep{molscat}\footnote{\url{http://www.giss.nasa.gov/tools/molscat}}
in which the extension to allow for the rotational structure of H$_2$
is not yet implemented. Hence, the present calculations were
restricted to collisions between NH$_3$ and p-H$_2$~($J=0)$. Extension
to p-H$_2$($J=0, 2$) and o-H$_2$($J=1$) is under way and is further
discussed below.

All calculations were performed at the close-coupling (CC) level.
Inversion doubling was neglected and the inversion-tunneling
wavefunction was simply taken as a linear combination of two delta
functions centered at the equilibrium position
\citep{green76,davis78}. We actually tested this approximation on the
NH$_3-$He system by taking the inversion coordinate explicitly into
account, as done previously by \citet{davis81}. To this aim, we
employed the high quality NH$_3-$He PES of \citet{hodges01}, which
does include the inversion dependence of the interaction. The
inversion motion was found to have a negligible effect (less than
10~\%) on the rigid-body interaction potential and on the cross
sections \citep{Scifoni07}, as was concluded by \citet{davis81} from a
lower quality potential. We note that \citet{sanden92} obtained a
comparable result for the NH$_3-$Ar interaction. A similar conclusion
is therefore expected for the NH$_3-$H$_2$ interaction, although the
inversion dependence of this PES is yet not known.

We adopted the rotational constants $A=B=9.944116$~cm$^{-1}$ and
$C=6.228522$~cm$^{-1}$. The reduced mass of the system is
1.802289~a.m.u. As the ortho- and para-levels of ammonia do not
interconvert in inelastic collisions, these were treated separately.
The coupled-channel equations were integrated using the modified
log-derivative propagator of \citet{manolopoulos86}. The radial
propagation used a stepsize parameter STEPS=10 except at total
energies below 30~cm$^{-1}$ where STEPS was increased up to 300 to
constrain the step length of the integrator below $\sim$0.1~Bohrs.
Other propagation parameters were taken as the \textsc{MOLSCAT}
default values. Calculations were performed for collision energies
between $\sim 0.1$~cm$^{-1}$ and 700~cm$^{-1}$. The energy grid was
adjusted to reproduce all the details of the resonances, with an
energy step of 0.2~cm$^{-1}$ up to total energies of 150~cm$^{-1}$ and
0.5~cm$^{-1}$ from 150 to 300~cm$^{-1}$. All calculations also
included several energetically closed channels to ensure that cross
sections were converged to within $7-8$\% for all transitions
involving $J\leq 3$. Thus, at the highest investigated energies, the
basis set incorporated all target states with $J\leq 11$ and 12 for o-
and p-NH$_3$, respectively.

\begin{figure}
  \centering 
  \includegraphics*[width=\columnwidth]{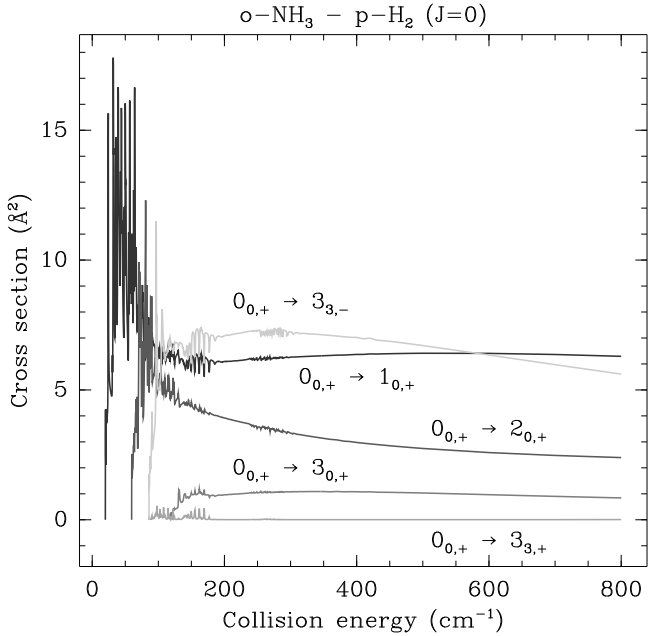}
  \caption{Cross sections for transitions out of the $0_{0,+}$
    rotational level of o-NH$_3$ into levels $J_{K,\epsilon}$ ($J\leq
    3$) as a function of collision energy.}
  \label{fig:xs}
\end{figure}

Excitation cross sections for o-NH$_3$ are presented in
Fig.~\ref{fig:xs} for rotation-inversion transitions out of the ground
state of o-NH$_3$. Prominent resonant features are observed in this
plot. These are caused by both Feshbach and shape type resonances. It
should be noted that only Feshbach type resonances are observed in
NH$_3-$He collisions \citep[e.g.][]{machin05}. This difference between
He and H$_2$($J=0$) reflects the deeper potential well of the
NH$_3-$H$_2$ PES, as discussed in Sect.~\ref{abinitio}. Resonances are
found to significantly increase the cross sections at low energy and,
therefore, the rate coefficients at low temperature. For example, at
10~K, the NH$_3-$H$_2$ rate coefficient for the ground-state
transition of o-NH$_3$ $1_{0,+} \rightarrow 0_{0,+}$ is a factor of 20
larger than the NH$_3-$He rate of \citet{machin05}. We note, however,
that this factor reduces to 2.5 for the transition $2_{0,+}
\rightarrow 0_{0,+}$ at the same temperature. \citet{willey02} also
reported significant differences (up to a factor of 4) between p-H$_2$
and He broadening cross sections. It is also noticed that the cross
section for the $0_{0,+} \rightarrow 3_{3,+}$ transition is much lower
than for $0_{0,+} \rightarrow 3_{3,-}$. This propensity rule was
already observed in earlier calculations \citep{offer90,rist93} but,
interestingly, it was found to be considerably weakened for collisions
with o-H$_2$~($J=1$) and it was observed experimentally only to a
slight extent \citep{schleipen93}. On the other hand, it was found to
be preserved when including the $J=2$ state in the p-H$_2$ basis set
\citep{offer90,rist93}. This inclusion was also found to change the
absolute values of the cross sections, at a few selected energies, by
up to a factor of 3 \citep{offer90,rist93}. Its effect on the average
cross sections and rate coefficients is however expected to be
moderate, typically 20$-$30~\%. This was indeed checked in the case of
ND$_2$H$-$H$_2$ calculations employing the present PES (Scifoni et
al., in preparation). As a result, the rate coefficients presented
below are expected to be accurate within typically 30~\%.

Cross sections were integrated over Maxwell-Boltzmann distributions of
collisional velocities and collisional rate coefficients were obtained
in the range 5$-$100~K for all transitions involving ammonia levels
with $J\leq 3$ (the lowest levels with $J=4$ lie at 177 and 237~K
above the ground states of p- and o-NH$_3$, respectively).  Higher
levels and temperatures were not investigated in the present work
because collisional rates with o-H$_2$ are required in models
considering temperatures above 100~K. These collisional rate
coefficients are made available in the
BASECOL\footnote{\url{http://www.obspm.fr/basecol/}} and
LAMBA\footnote{\url{http://www.strw.leidenuniv.nl/~moldata/}}
databases as well as at the
CDS\footnote{\url{http://cdsweb.u-strasbg.fr/}}.

\begin{figure}
  \centering 
  \includegraphics*[width=\columnwidth]{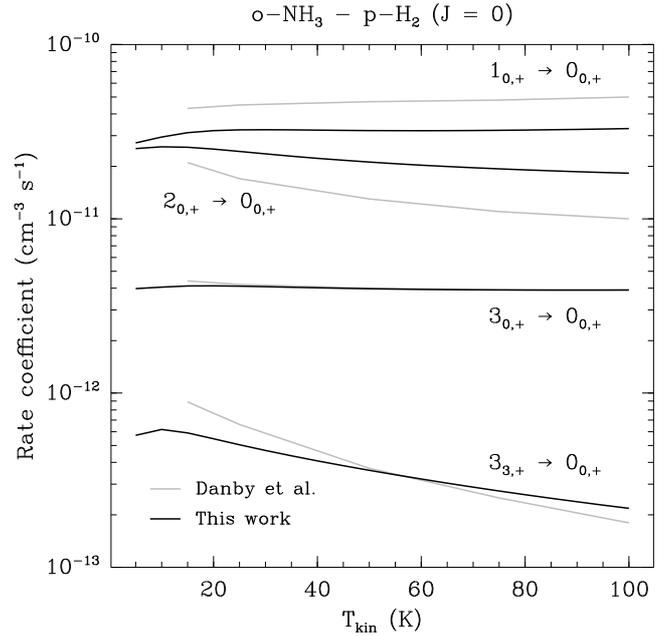}
  \caption{Comparison between rate coefficients for collisional
    de-excitation of o-NH$_3$ by p-H$_2$ ($J=0$) of \citet{danby86}
    and those from this work, as a function of temperature. Only
    transitions towards the ground state $0_{0,+}$ are displayed.}
  \label{fig:o-rate}
\end{figure}

In Fig.~\ref{fig:o-rate}, downward rate coefficients are presented for
rotation-inversion transitions towards the ground state. The resonant
features are found to be completely washed out by the thermal average.
The present results are compared with the data of \citet{Danby88}. As
expected from the comparison of the expansion angular coefficients
(see Fig.~\ref{fig:expcoef}), the new rates agree within a factor of 2
with those of \cite{Danby88}. We see, however, that there is no
particular trend, although the present rates are generally larger.

\begin{figure}
  \centering 
  \includegraphics*[width=\columnwidth]{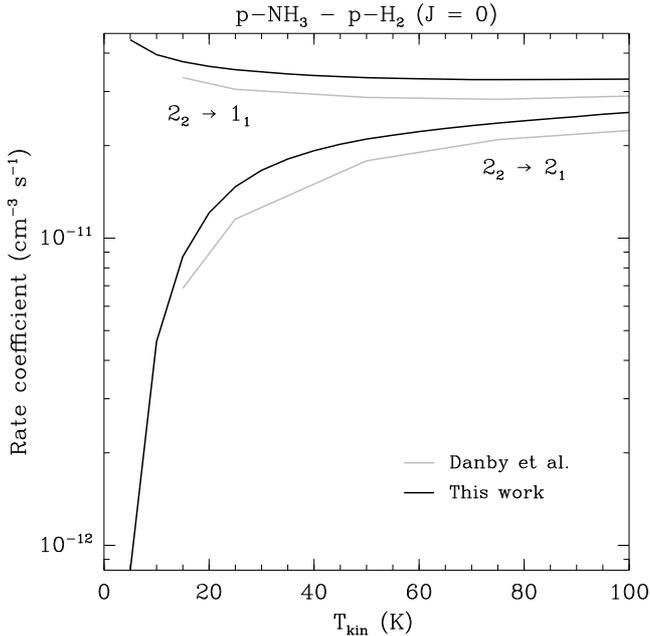}
  \caption{Comparison between the rate coefficients for collisional
    excitation and de-excitation of p-NH$_3$ by p-H$_2$~$(J=0)$ of
    \citet{danby86} and those from this work, as a function of
    temperature.  These rates have been averaged over the symmetry
    index $\epsilon$}
  \label{fig:p-rate}
\end{figure}

In Fig.~\ref{fig:p-rate}, we show the symmetry-averaged rates $C
\left( 2_{2} \rightarrow 2_{1} \right)$ and $C \left( 2_{2}
  \rightarrow 1_{1} \right)$ that appears in Eq.~(\ref{eq:1}), as a
function of the temperature. The new rates are found to be larger than
those of \citet{Danby88} by typically 15\% but to follow closely the
same temperature dependence.  As a result, minor modifications of the
ammonia thermometer are expected, as shown below.

\section{Non-LTE excitation computations}
\label{sec:non-lte-excitation}

In order to estimate the effect of the new collisional rates on the
calibration of the ammonia thermometer, we have computed the
excitation of both o-NH$_{3}$ and p-NH$_{3}$ using the large velocity
gradient code of \citet{vanderTak07}. We have used the o-NH$_{3}$ and
p-NH$_{3}$ collisional rates with p-H$_{2}$ presented in the previous
section, as well as those from \citet{Danby88}, for comparison. The
latter were taken from the LAMBA database
\citep{Schoier05a}. Collision between NH$_{3}$ and He were neglected,
because, in addition to H$_{2}$ being more abundant than He by a
factor of 5, NH$_{3}$-H$_{2}$ collision rates are typically a factor
of 3 larger than the NH$_{3}$-He rates \citep{machin05}.  Energy
levels, statistical weights and Einstein coefficients were taken from
the JPL database for molecular spectroscopy \citep{Pickett98}. For the
calculations using \citet{Danby88} collisional rates, the first 24
levels of o-NH$_{3}$ and the first 17 levels of p-NH$_{3}$ were
considered (corresponding to energy levels up to 416 and 297
cm$^{-1}$, respectively). For the calculations using the new rates,
only the first 6 levels of o-NH$_{3}$ and the first 10 levels of
p-NH$_{3}$ were considered (up to 118 and 115 cm$^{-1}$,
respectively). In both cases, we have neglected the hyperfine
structure of the molecule, i.e. we have considered that each hyperfine
level within a given inversion level corresponds to the same energy
level. While this hypothesis will lead to an overestimate of the line
opacity for optical depths greater than a few, it is valid if the line
is optically thin \citep{Daniel06}. We have therefore chosen a column
density to velocity gradient ratio that is large enough for this
approximation to be valid ($N(\mathrm{NH_{3}})/(dv/dr) = 10^{-4} \,
\mathrm{cm}^{-3} / (\mathrm{km} \, \mathrm{s}^{-1} \,
\mathrm{pc}^{-1})$; the same value adopted by \citet{Walmsley83}).

\begin{figure}
  \centering
  \includegraphics[width=\columnwidth]{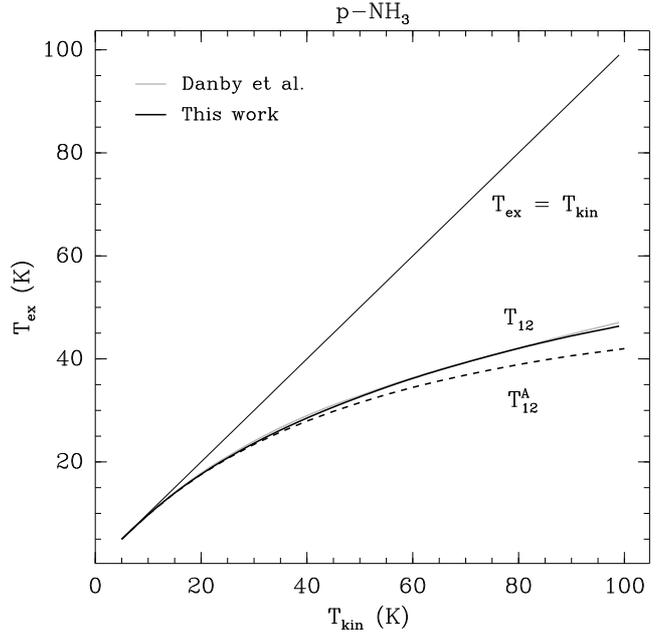}
  \caption{Rotational temperature between the $1_1$ and $2_2$
    metastable levels of p-NH$_{3}$ computed using the collisional
    rates of \citet[]{Danby88} and those from this paper, as a
    function of the kinetic temperature. The dashed curve show the
    rotational temperature computed from Eq.~\ref{eq:1}.}
  \label{fig:t12-tkin}
\end{figure}

Fig. \ref{fig:t12-tkin} shows the excitation temperature between the
$1_1$ and $2_2$ metastable levels computed using both sets of
collisional rates and as a function of the kinetic temperature. This
excitation temperature is obtained by summing the populations of the
$\epsilon = +1$ and $\epsilon = -1$ within each rotational state. A
p-H$_{2}$ density of 10$^{4}$ cm$^{-3}$ was assumed. On this figure,
we also show the excitation temperature computed from Eq.
(\ref{eq:1}), i.e. assuming that only the first three rotational
levels are populated. As seen on this figure, the excitation
temperature computed using the rates of \cite{Danby88} and the one
computed using the rates presented in this paper agree extremely well;
both values differ by less than 2\%. We also notice that for kinetic
temperatures lower than 20~K, the excitation temperature is well
approximated by Eq.~(\ref{eq:1}), but it underestimates it at larger
temperature, because higher energy levels start to become populated.
The good agreement between the kinetic temperature obtained using the
rates of \cite{Danby88} and those presented in this paper can be
simply understood by examination of Eq.~(\ref{eq:1}). In this
approximation, the excitation temperature depends on $\ln (1 +
C_{21}/C_{23})$. Although the rates computed in this paper differ by
$\sim$15\% with respect to those of \cite{Danby88}, their ratio (and a
fortiori the logarithm of their ratio) differ much less. Therefore,
the relation between the excitation temperature and the kinetic
temperature -- or in other words the calibration of the ammonia
thermometer -- appears to be robust.

\begin{figure}
  \centering
  \includegraphics[width=\columnwidth]{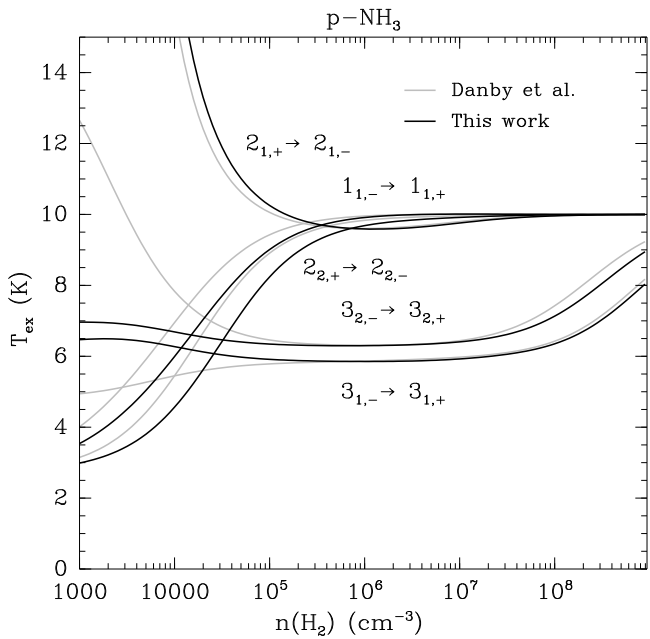}
  \caption{Excitation temperature of the p-NH$_{3}$ inversion
    transitions computed using the collisional rates of
    \citet[]{Danby88} and those from this paper, as a function of the
    H$_{2}$ density. A kinetic temperature of 10~K is assumed.}
  \label{fig:tex-pnh3-nh2}
\end{figure}

\begin{figure}
  \centering
  \includegraphics[width=\columnwidth]{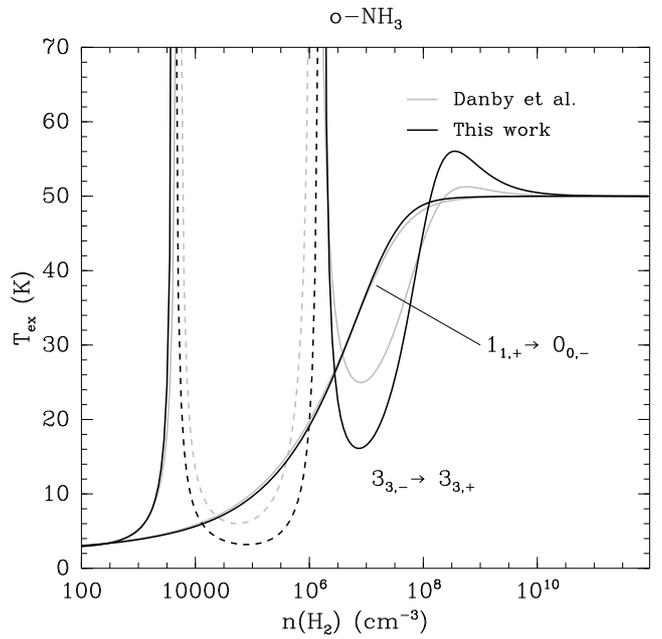}
  \caption{Same as in Fig. \ref{fig:tex-pnh3-nh2} for the o-NH$_{3}$
    $3_{3,-}-3_{3,+}$ inversion line and for a kinetic temperature of
    50~K. The dashed lines indicate negative excitation temperatures.}
  \label{fig:tex-onh3-nh2}
\end{figure}

\begin{table*}
  \caption{
    Frequencies, upper level energies and critical densities of the
    p-NH$_3$ and o-NH$_3$ inversion and rotation-inversion transitions
    considered in this paper. Spectroscopic data are from the JPL
    catalog. Energies are relative to the fundamental rotation-inversion
    level of each specie, i.e. $1_{1,+}$ for p-NH$_{3}$ and $0_{0,+}$ for
    o-NH$_{3}$. The critical densities are given for a kinetic temperature
    of 10~K. They are computed using Eq.~\ref{eq:4}, and thus refer to the
    upper of level of the transition.
    \label{tab:freq}}
  \centering
  \begin{tabular}{l c c c c}
     \hline
     \hline
     Species & Transition & $\nu$  & $E_{up}$ & $n_\mathrm{crit}$ \\
     & $J_{K,\epsilon} - J'_{K',\epsilon'}$ & (GHz) & (K) & (cm$^{-3}$) \\
     \hline
     p-NH$_{3}$ & $1_{1,-} \rightarrow 1_{1,+}$ &  23.694496 &   1.1 & $3.90 \times 10^{3}$ \\
     p-NH$_{3}$ & $2_{2,+} \rightarrow 2_{2,-}$ &  23.722633 &  42.3 & $3.08 \times 10^{3}$ \\
     p-NH$_{3}$ & $2_{1,+} \rightarrow 2_{1,-}$ &  23.098819 &  58.3 & $1.44 \times 10^{8}$ \\
     p-NH$_{3}$ & $3_{2,-} \rightarrow 3_{2,+}$ &  22.834185 & 128.1 & $3.01 \times 10^{8}$ \\
     p-NH$_{3}$ & $3_{1,-} \rightarrow 3_{1,+}$ &  22.234506 & 144.0 & $5.41 \times 10^{8}$ \\
     o-NH$_{3}$ & $3_{3,-} \rightarrow 3_{3,+}$ &  23.870129 & 123.6 & $2.63 \times 10^{3}$ \\
     o-NH$_{3}$ & $1_{1,+} \rightarrow 0_{0,+}$ & 572.498068 &  27.5 & $5.45 \times 10^{7}$ \\
     \hline
   \end{tabular}
 \end{table*}

Fig. \ref{fig:tex-pnh3-nh2} shows excitation temperature of several
p-NH$_{3}$ inversion transitions, as a function of the density, for a
kinetic temperature of 10~K and the same column density and line
velocity than in Fig. \ref{fig:t12-tkin}. Spectroscopic data and
critical densities are given in Table~\ref{tab:freq}. For a
multi-level system, the critical density can be defined (in the
optically thin case) as the density at which the sum of the
collisional de-excitation rates out of a given level is equal to the
sum of the spontaneous radiative de-excitation rates:

\begin{equation}
  n_\mathrm{crit}(T)=\frac{\sum_{J'_{K',\epsilon'}}A(J_{K,\epsilon}\rightarrow
    J'_{K',\epsilon'})} {\sum_{J'_{K',\epsilon'}}C(J_{K,\epsilon}\rightarrow J'_{K',\epsilon'})(T)}.
  \label{eq:4}
\end{equation}

\noindent where the summation is done over the $J'_{K'\epsilon'}$
levels (with energies smaller than that of the $J_{K,\epsilon}$
considered). With this definition, the critical density refer to a
level, and not to a transition.

For densities lower than $10^{3}$~cm$^{-3}$, the excitation
temperature of the $1_{1,-} \rightarrow 1_{1,+}$ and $2_{2,+}
\rightarrow 2_{2,-}$ inversion transitions computed using the rates
presented here and those of \citeauthor{Danby88} show little
differences. For these densities, collisional de-excitation is
negligible, and the excitation temperature of these lines are close to
the background temperature (2.73~K). For densities much greater than
the critical density (i.e. $\gtrsim 10^{6}$~cm$^{-3}$), collisional
de-excitation dominates, and lines are essentially thermalized.  At
intermediate densities, the $1_{1,-} \rightarrow 1_{1,+}$ and $2_{2,+}
\rightarrow 2_{2,-}$ line excitation temperatures predicted using the
rates from this work is slightly lower than the one predicted using
those from \citeauthor{Danby88}.  This is because the de-excitation
rates from this work are smaller (by about a factor two) than those of
\citeauthor{Danby88} for these lines. As a consequence, the critical
densities of the corresponding levels is greater than previously
estimated, and the transition thermalize at greater densities. Larger
differences in the excitation temperatures of the $3_{2,-} \rightarrow
3_{2,+}$ and $3_{1,-} \rightarrow 3_{1,+}$ transitions -- for which
critical densities are a few $10^{8}$~cm$^{-3}$ -- are seen. For
example, at a density of $10^{3}$~cm$^{-3}$, the excitation
temperature of the $3_{2,-} \rightarrow 3_{2,+}$ transition computed
using the rates of \citeauthor{Danby88} is almost a factor two larger
than the one computed with the rates obtained here. From the observer
point of view, this has no consequences because the energy of the
upper level of the transition is 123.6~K. At low densities, for the
kinetic temperature considered here, the fractional population of this
level is extremely small, and the predicted antenna temperature is
essentially zero.

Fig. \ref{fig:tex-onh3-nh2} shows the excitation temperature of the
o-NH$_{3}$ $3_{3,-} \rightarrow 3_{3,+}$ inversion transition as a
function of the density, for the same column density to velocity
gradient ratio than in Fig. \ref{fig:tex-pnh3-nh2}, but -- since the
upper level of the transition lies at $\sim 124$~K above the ground
level of o-NH$_{3}$ -- a kinetic temperature of 50~K \footnote{The
  computations with the rates from the present work do not include the
  levels above $J = 3$. To make sure that these levels can be
  neglected at a kinetic temperature of 50~K, we have computed the
  excitation temperature for the transitions shown on Fig.~7 using
  Dandy's rates, but without considering the levels above $J=3$. These
  were found to be quasi-identical to those computed when the levels
  above $J=3$ are considered.}  The behavior of the excitation
temperature is similar to that of p-NH$_{3}$ inversion lines; at low
density, it is close to the background temperature, while it is
thermalized at densities greater than $10^{6}$~cm$^{-3}$. We predict,
in agreement with \citet{Walmsley83}, a population inversion for
densities ranging between $\sim 4 \times 10^{3}$ and $\sim 6 \times
10^{5}$~cm$^{-3}$ (note that the range in which the inversion occurs
is slightly different for the two collisional rate sets). This
population inversion was first predicted by \citet{Walmsley83}, and
has been studied in detail by \citet{Flower90}. The corresponding
maser transition has been since observed in several star forming
regions, e.g. NGC~6334I \citep{Beuther07}. As explained by
\citet{Walmsley83}, the lower level of the transition ($3_{3,+}$) is
de-populated by collisions to excited levels in the $K = 0$ ladder,
while it is populated by radiative transitions from the $3_{3,-}$
level. For densities greater than $\sim 4 \times 10^{3}$~cm$^{-3}$,
the collisional de-population rate is greater than the radiative
population rate of the lower level, and the inversion occurs. For
densities greater than $\sim 6 \times 10^{5}$~cm$^{-3}$, excited
levels in the $K = 0$ ladder start to become populated and populate
the $3_{3,+}$ level collisionally. This limits the maser gain to
moderate values; for a density of $1 \times 10^{5}$~cm$^{-3}$, we
predict a negative opacity of only $\tau \sim - 2$.

Fig.~\ref{fig:tex-onh3-nh2} also shows the excitation temperature of
the $1_{1,+} \rightarrow 0_{0,+}$ rotation-inversion transition of
o-NH$_{3}$. This line, at a frequency of $\sim 572.5$~GHz, was first
detected towards OMC-1 with the \emph{Kuiper Airborne Observatory}
\citep{Keene83}. It was also detected towards $\rho$-Oph~A with
\emph{Odin} space telescope \citep{Liseau03}, and it will be soon
observable with the \emph{Heterodyne Instrument for the Far Infrared}
(HIFI) on board the \emph{Herschel Space Observatory}. This line is
found to thermalize at densities of $\sim 10^{9}$~cm$^{-3}$. Once
again, little differences between the excitation temperatures computed
using the rates of Danby et al. and those from this work are seen. No
significant difference were found between the excitation temperatures
of the other rotation-inversion transitions that will be observable
with HIFI either.

\section{Conclusions}
\label{sec:conclusions}

We have presented new collisional excitation rates of p-NH$_{3}$ and
o-NH$_{3}$ with p-H$_{2}$($J=0$). With respect to older computations
from \cite{Danby88}, the present rates were found to agree within a
factor of 2. In order to investigate the effect of the new rates on
the excitation of o-NH$_{3}$ and p-NH$_{3}$, we have computed the
excitation of these species under physical conditions that are typical
of dense molecular clouds, prestellar cores, as well as the outer
envelopes of embedded protostars, using an LVG code. We found that the
excitation temperature between the $1_1$ and $2_2$ levels computed
using the new rates is almost identical to that computed using older
rates at the low temperatures considered here ($\leq 50$~K). Thus, the
calibration of the ammonia thermometer appears robust.  The effect of
the new rates on the inversion transitions (at cm wavelenght) or the
rotation-inversion transitions that will be observable with
Herschel-HIFI are also found to be small.

Future works include extension of the present calculations to
p-H$_2$~$(J=0,2)$ and o-H$_2$~$(J=1)$ as well as to higher ammonia
levels and kinetic temperatures. Comparisons with double resonance
\citep{daly70}, crossed beam \citep{schleipen93} and pressure
broadening \citep{willey02} experiments will be also investigated,
with the objective to establish the predictive ability of the present
PES and to distinguish between the predictions of the available PES.

\section*{Acknowledgments}
The authors wish to acknowledge their friend and colleague Pierre
Valiron who initiated the work that lead to this paper. Pierre passed
away on August, 31th 2008, and he is deeply missed. We also thanks
Evelyne Roueff for her critical reading of this manuscript. This
research was supported by the CNRS national program ``Physique et
Chimie du Milieu Interstellaire'' and by the FP6 Research Training
Network ``Molecular Universe'' (contract number MRTN-CT-2004-512302).

\bibliographystyle{mn2e}
\bibliography{bibliography}

\label{lastpage}

\end{document}